\documentstyle[aps,twocolumn]{revtex}

\def\begineq{\begin{equation}}
\def\endeq{\end{equation}}
\def\be{\begin{equation}}
\def\ee{\end{equation}}

\begin{document}
\bibliographystyle{prsty}

\title{Sonoluminescing air bubbles rectify argon}
\author{Detlef Lohse $^1$, 
Michael P. Brenner $^2$, Todd F. Dupont $^3$,Sascha Hilgenfeldt $^1$,
and Blaine Johnston $^4$} 
\address{$^1$
Fachbereich Physik der Universit\"at Marburg,
Renthof 6, 35032 Marburg, Germany,\\
$^2$
Department of Mathematics,
Massachusetts Institute of Technology, Cambridge, MA 02139,\\
$^3$ Department of Computer Science, 
University of Chicago, Chicago, IL 60637\\
$^4$ Department of Physics, 
University of Chicago, Chicago, IL 60637\\
}
  
\date{\today}

\maketitle
\date{\today}

\begin{abstract}
The dynamics of
single bubble sonoluminescence (SBSL) strongly depends
on the percentage of inert gas within the bubble.
We propose a theory for this dependence, based on
a combination of principles from sonochemistry and hydrodynamic
stability.
The nitrogen and oxygen dissociation and 
subsequent reaction to water soluble gases
implies that
strongly forced air bubbles eventually consist of pure argon.
Thus it is the partial argon (or any other inert gas)
 pressure which is relevant for
stability. 
The theory  provides quantitative explanations for many aspects
of SBSL.
\end{abstract}
\pacs{PACS: 78.60.Ma, 82.40.We, 42.65.Re, 43.25.+y, 82.40.-g}


Recent experiments 
 \cite{cru94}
revealed that a single gas bubble levitated in a strong acoustic field
$P(t)=P_a \cos \omega t$ can emit picosecond bursts of light,
a phenomenon called single bubble sonoluminescence (SBSL).
The phase and intensity
of the light can be stable for hours.  SBSL shows
a sensitive dependence on
the forcing pressure, the concentration of the dissolved gas, and the 
liquid temperature, among other parameters.

Particularly puzzling is the dependence on the
type of the gas dissolved in the
liquid.
Hiller {\it et al.} \cite{hil94}
demonstrated that stable SBSL
does not occur with pure nitrogen, oxygen or 
nitrogen--oxygen
mixtures.
But a critical concentration 
of argon gives stable
SBSL.  
This paper presents an explanation for
this dependence on the type of gas,
based on combining principles
from sonochemistry \cite{sch36,ver88} with hydrodynamic stability
\cite{hil96}.

An important clue comes from experiments \cite{bar95,loe95} 
showing that the 
range of gas concentrations for stable SBSL in pure
argon bubbles differs from that for air bubbles by two orders of
magnitude.
Stable SBSL in argon requires strong degassing
of the liquid down to  the tiny 
gas pressure $p_\infty = 0.004 P_0$, which by Henry's law
corresponds to a concentration $c_\infty = 0.004 c_0$, where $P_0=1$atm 
and $c_0$ is the saturation concentration. 
In contrast to argon, SBSL with air only requires degassing down to 
$p_\infty / P_0 = 0.1-0.4$ \cite{gai90,bar95}.   
L\"ofstedt {\it et al.} \cite{loe95} estimated
that diffusive equilibrium of the bubble with the surrounding
dissolved gas requires  $p_\infty/P_0 \sim 10^{-3}$, suggesting
agreement with the experiments for argon,
but not air. Indeed,
detailed hydrodynamic stability calculations \cite{hil96}
show quantitative agreement with argon data.
The strong discrepancy led L\"ofstedt
{\it et al.} to the
conjecture that there is
an ``as yet unidentified mass ejection
mechanism'' in air bubbles which ``is the key to SL in a single bubble''.

We suggest 
that this mechanism is chemical.
The importance of
chemical reactions has long been recognized in
multibubble sonoluminescence (MBSL) in
transient cavitation clouds \cite{mar85,ver88},
since 
the pioneering work of
Schultes and Gohr \cite{sch36} found that aqueous solutions of nitrogen
produced nitric and nitrous acids when subjected to ultrasound.
High temperatures generated by 
the bubble collapse are beyond the
dissociation temperature of oxygen and nitrogen ($\approx 9000$K), leading
to the formation of O and N radicals which react with the H and O 
radicals formed from the dissociation of 
water vapor. Rearrangement of the radicals will lead to the formation
of NO, OH, NH, which eventually dissolve in water to form HNO$_2$ and
HNO$_3$, among other products. 

Based on fits of SBSL spectra \cite{ber95,hil92} and hydrodynamic
calculations \cite{loe93,bar94},
it is well accepted 
that internal bubble temperatures in SBSL are 
even higher than in MBSL. Therefore, the same reactions as in MBSL
will occur.
The reaction products (NO$_2$, NO, $\dots$) are
pressed into the surrounding liquid, and are not recollected during
the next bubble cycle, since their solubility 
 in water is enormous.
This chemical process deprives the gas in the bubble of its reactive
components.
Small amounts of N$_2$ and O$_2$ that diffuse into
the bubble during the
expansion react and their dissociation products are expelled back into
the surrounding liquids at the bubble collapse.
The only gases that can remain within a SBSL bubble over
many bubble cycles are those which at high temperatures
do not react with the liquid vapor,
i.e., inert gases. 
Hence, when air is dissolved in water, a strongly forced
bubble 
is almost completely 
filled with argon. This  
argon rectification happens in SBSL but not in MBSL because
it requires bubble stability over many oscillation cycles.

In the following, we first present qualitative consequences of
argon rectification, demonstrating that even at a crude
level it resolves central problems of
bubble stability.  Then we proceed to make the argument more precise 
through quantitative calculation.

If the bubble is filled essentially with argon, the hydrodynamic
stability of the bubble is determined by
the {\it partial} pressure
of argon $p_\infty^{Ar} = \xi_{l} 
p_\infty $,
where $\xi_{l}$ ($=0.01$ for air)
is the argon ratio of gas dissolved in the liquid.
This fact immediately resolves the hundredfold difference
between the amount of degassing necessary for air
versus argon: As mentioned
above,
hydrodynamic stability calculations \cite{hil96}
demonstrate that  stable sonoluminescence
for argon bubbles exists between $p_{lower} < 
p_\infty^{Ar}/P_0 < p_{upper} $,
with $[p_{lower},p_{upper}]$ depending on the
forcing pressure $P_a$ (figure \ref{figure1}). E.g. at $P_a=1.3$atm, 
$[p_{lower},p_{upper}]\approx [0.002,0.004]$.
Since the stability window for air bubbles is set
by the partial pressure of argon
$p_\infty^{Ar}=0.01p_\infty$, 
the total pressure
of the air mixture must be 
in the range
$[100p_{lower}, 100p_{upper}] $ for stable SL.
At $P_a=1.3$atm, this corresponds to
$0.2 < p_\infty /P_0 < 0.4$, in good agreement with
experiments.
One reason that air with its
$1\%$ argon is a particularly friendly
gas for SL experiments is that this
amount of degassing
is easily achieved. 

\begin{figure}[htb]
\setlength{\unitlength}{1.0cm}
\caption[]{
Phase diagram for pure argon bubbles in the
$p^{Ar}_\infty/P_0^{}$ versus $P_a/P_0$ parameter space,
from \cite{hil96}, but now with experimental data included.
Stable SL is only possible
in a very small window of argon concentration.
The experimental data points refer to observed stable
SL (filled symbols) or stable non-SL bubbles
(open symbols) and are extracted (using the present theory)
from refs.\ \cite{loe95} (diamonds) and \cite{hol96} (circles)
and show good agreement with the theory.
}
\label{figure1}
\end{figure}

At this simple level, the theory
makes several other explicit predictions:  
When
varying the percentage
of argon in N$_2$-Ar mixtures the
range of $p_\infty/P_0$ where SL is
stable should vary
like $p_{lower}/\xi_l < p_\infty/P_0 < p_{upper}/\xi_l$,
with the upper threshold, the lower threshold, and the
range increasing
with decreasing
argon fraction $\xi_l$.
Another consequence 
is that there is an
argon ratio
($\xi_{l}
\approx 0.003$ at $P_a=1.3$atm) for which
stable SL should be possible {\it without degassing}.  We caution that
to achieve this
it is necessary to
rid the liquid of impurities to avoid
spontaneous cavitation.

Also the different character of the transition to SBSL observed
in air bubbles and in argon bubbles 
can be explained by our theory:
Hiller {\it et al.} \cite{hil94} showed that
for pure argon bubbles, the bubble
mass
increases smoothly and monotonically upon increasing the
forcing pressure.  In contrast Barber {\it et al.}
\cite{bar95} found that the
transition to SBSL in air bubbles causes an abrupt decrease in the bubble
mass.
The difference between these two experiments is the presence or
absence of sonochemical reactions:
In air bubbles
an abrupt behavior occurs when
the forcing pressure
corresponds to the onset of the dissociation reaction.  Below this
threshold, the bubble contains air, and
the ambient radius is determined by the diffusive stability of the mixture.
Above the transition, when the molecular gases dissociate, the
equilibrium radius is set by diffusive stability of
pure argon bubbles resulting in a much different ambient
radius \cite{hil96}.
The transition towards SL is smooth for pure argon because
a dissociation mechanism is absent.

We now turn to a quantitative calculation of hydrodynamic
stability
for gas mixtures with
chemical reactions.
The bubble radius $R(t)$ 
is well described by the Rayleigh-Plesset equation
\cite{loe93}, which we take to have the 
same form and parameters as in our earlier work \cite{hil96}. 
The internal bubble pressure $p(t)$
is assumed to obey a van der Waals equation
of state. 
Now consider a bubble in water
containing a mixture
of a reactive gas (taken to be N$_2$) and an inert gas, Ar.
The total number of moles of gas in the bubble is 
$N_{tot} = 4\pi R_0^3 P_0 / (3G\Theta_0 ) = N_{N_2} + N_{Ar}$, where 
$\Theta_0 = 273$K is the normal temperature
and $G=8.3143J/(molK)$ is the gas constant.
The  argon fraction {\it in} the bubble is
$\xi_b = N_{Ar}/N_{tot}$, that of nitrogen $1-\xi_b = N_{N_2}/N_{tot}$. 
If $c^{Ar}(r,t)$ and $c^{N_2}(r,t)$ are the 
mass-per-volume concentration fields  of
Ar and N$_2$ in the liquid, respectively,
the rate of change of the
number of molecules 
of N$_2$ and Ar in the bubble is given by
\begin{eqnarray}
\dot{N}_{Ar} &=& {{4\pi R^2 D_{Ar}} \over{\mu_{Ar}}} \partial_r c^{Ar}\vert_{r=R}\label{ar}\\
\dot{N}_{N_2}&=&  {{4\pi R^2 D_{N_2}} \over{\mu_{N_2}}} \partial_r c^{N_2}\vert_{r=R}
- A N_{N_2} 
 \exp{\left(- {\Theta^*\over \Theta}\right)}\label{n2}.
\end{eqnarray}
Here, $D_{Ar}$, $D_{N_2}$,
$\mu_{Ar}$ and $\mu_{N_2}$ are the respective
diffusion constants and molecular masses.
The concentration fields obey an advection diffusion equation,
whose boundary conditions are set by the external concentrations
$c^\alpha (\infty , t) = c_\infty^\alpha = p_\infty^\alpha
c_0^\alpha/P_0$ (Henry's law)
and by the partial gas
pressures $p^\alpha (t) $ in the bubble $c^\alpha (R(t),t) =
p^\alpha (R(t)) c_0^\alpha / P_0$, $\alpha =$  Ar,  N$_2$.
The second term in (\ref{n2}) represents the bubble's nitrogen loss
by chemical reactions.
The reaction rate depends on the temperature $\Theta(t)$ in the
bubble.
For simplicity,
we assume that the reactions follow an Arrhenius law, with
empirical parameters  appropriate for nitrogen dissociation
(ref.\ \cite{ber96}):
$A=6 \cdot 10^{19} (\Theta_0 / \Theta)^{2.5}
(\rho_0/\mu_{N_2})(R_0/R)^3cm^3/(mols)$
giving the timescale
of the reaction;
$\Theta^* = 113000K$ is the activation temperature and
 $\rho_0$
the equilibrium gas density.
This reaction law is rather crude, as it
neglects backward reactions as well as the kinetics of the 
expulsion of reaction products; however, 
it is sufficient for this demonstrative calculation.

The diffusive mass flux into the bubble can
be calculated explicitly
using the fact that the diffusive timescale is much
slower than the bubble oscillation period $T$ \cite{ell69,fyr94,loe95,hil96}.
This
reduces the
diffusional problem to
the calculation of weighted averages 
of the form
$\left< f \right>_i = {\int_0^T f(t) R^i(t) dt /
\int_0^T R^i(t) dt} $,
with the mass flux
proportional to
$p_\infty - \left< p \right>_4 $ for a pure gas \cite{fyr94}.
Applying the same approximation to equations (\ref{ar}) and (\ref{n2})
gives
\begin{eqnarray}
{\Delta N_{Ar} \over T} &\approx & {4\pi R_{m} D_{Ar} c_0^{Ar} 
\over \mu_{Ar}  P_0 } 
\left( 
p_\infty^{Ar} - \xi_b
\left< p \right>_4
\right)
\label{ndot_ar}
\\
{\Delta N_{N_2}\over T} &\approx & {4\pi R_{m} D_{N_2} c_0^{N_2}
\over \mu_{N_2} P_0 } 
\left( 
p_\infty^{N_2} -  (1-\xi_b)
\left< p \right>_4 
\right)
\nonumber \\ &-&
 {N_{N_2}} \left< A
\exp{\left(- {\Theta^* /  \Theta}\right)}\right>_0,
\label{ndot_n2}
\end{eqnarray}
$R_m=\hbox{max}_t R(t)$. 
To close the equations, we need a model
for the temperature dependence $\Theta(t)$.
The actual temperature dependence is determined
by complicated nonlinear processes operating
during the collapse.
As a simple model, we take the temperature to be
uniform within the bubble, and use the
polytropic law
$
\Theta(t) = \Theta_0 \left( {( R_0^3 -h^3 )/
(R^3(t) - h^3) }
\right)^{\gamma - 1 }
$
with $h$ the van der Waals hard core radius and $\gamma$
the polytropic exponent.  
The value of $\gamma$ depends on the rate of
heat transport from the bubble, which
is characterized by
the P\'eclet number $Pe = \dot R
R_0^2/
(R \kappa )$, where
 $\kappa$ is the thermal diffusivity of the gas.
During the bubble expansion, $Pe \ll 1$  (isothermal behavior, 
$\gamma=1$);
during the
bubble collapse, $Pe \gg 1$ (adiabatic compression, $\gamma=5/3$).
Since both of these regimes occur during a single bubble cycle,
it is necessary to use a model which interpolates between them.
For definiteness, we follow
Prosperetti
\cite{ple77}
and use his calculated
$\gamma (Pe)$.
We emphasize that although this treatment of the chemical reactions
is crude, the central results discussed below are robust.

\begin{figure}[htb]
\setlength{\unitlength}{1.0cm}
\caption[]{
Phase diagram for air at $p_\infty/P_0=0.2$ in the $R_0$ - $P_a$ space.
The arrows
denote whether the ambient radius grows or shrinks at this parameter
value.  Curve A denotes the  
equilibrium for an air bubble, on curve C the bubble only contains 
argon.   The intermediate curve B
necessarily exists because of the topology of the diagram, and represents
an additional stable equilibrium.  Above and right of
the thin line, the gas temperature
exceeds the nitrogen dissociation threshold of about $9000$K.
}
\label{r0vspa}
\end{figure}

With these approximations, the
equilibrium states
(satisfying $\Delta N_{Ar} =\Delta N_{N_2} = 0$)
can be computed as a function of
($N_{Ar}, N_{N_2}$), or equivalently as a function of ($\xi_b ,
R_0$).
Fig.\ \ref{r0vspa}
shows the equilibrium radii $R_0^*$ 
in the $R_0-P_a$ plane for air at $p_\infty/P_0=0.2$.
For small forcing the
temperatures
are not high enough to initiate chemical reactions, so
that the equilibrium curve corresponds to the classical prediction
by Eller and Flynn \cite{ell69} for this gas concentration.
This equilibrium is unstable: The bubble either shrinks or
grows by rectified diffusion; 
experiments \cite{bar94,bar95,gai90} show that a growing bubble
eventually
runs into a shape instability where
microbubbles pinch off
and make the bubble dance because of the recoil
\cite{hil96}. 
In the opposite limit of high forcing (curve C),
the reactions
burn off all the N$_2$, so that
the bubble contains pure argon; this equilibrium corresponds to the
(stable) Eller-Flynn equilibrium
at the argon partial pressure $p_\infty^{Ar}/P_0 = 0.01 p_\infty/P_0=0.002$.

Figure \ref{r0vspa} displays a regime of shrinking bubbles at high forcing
pressures (left of curve C) and an adjacent region of growing bubbles
(right of curve A).
This necessitates the existence of an additional
equilibrium at intermediate forcing pressures,
curve B in figure \ref{r0vspa}, for which growth by rectified diffusion
and loss by reactions balance.  This 
additional equilibrium occurs
close to the point of nitrogen dissociation,
and turns out to be stable; the argon fraction 
$\xi_b^*$ for this equilibrium is
slightly larger than the fraction $\xi_l$ in the liquid (for
not too strong forcing). --
The only feature of Fig.\ 2 that depends on the details of temperature
and chemical reactions is the exact position of the nitrogen dissociation
threshold and thus the exact position of curve B.

\begin{figure}[htb]
\setlength{\unitlength}{1.0cm}
\caption[]{
The equilibrium fraction
$\xi_b^*$ of argon in the bubble
as a function of $P_a$ and $R_0$.
}
\label{xi_fig}
\end{figure}

Figure \ref{xi_fig}
plots the 
equilibrium composition $\xi_b^*$,
given by $\Delta N_{Ar}
/\Delta N_{N_2} = \xi_b^* / (1-\xi_b^* ) $.
Weakly forced bubbles
have $\xi_b^* \approx \xi_l$, thus $p_\infty/P_0 = 0.20$ 
is relevant for stability.
Strongly forced bubbles have $\xi_b^*
\approx 1 \gg \xi_l$, thus $p_\infty^{Ar} /P_0= 0.002$ is the relevant
quantity.
The transition between these regimes is abrupt, and occurs when the bubble
temperature surpasses the dissociation temperature ($\approx 9000K$ for N$_2$).

What happens for even lower argon concentration
 $\xi_l < 0.01$ in the dissolved
gas? For these low concentrations 
the equilibria curves A and B hardly
depend on $\xi_l$. This holds even in the limit $\xi_l \to 0$ of pure nitrogen
bubbles. Therefore, 
our theory predicts that there is a parameter regime of forcing
pressures
where stable N$_2$ bubbles exist.
The equilibrium curve
C, of course, does depend on $\xi_l$ and for decreasing argon concentration
$p_\infty^{Ar} = \xi_l p_\infty$ it moves further and further to the right,
allowing diffusively stable SL bubbles only for larger and larger forcing. 
Finally, spherical instability will destroy these bubbles \cite{hil96}.

Having calculated the above
phase diagrams, we 
come back to a comparison 
to experiments. For an air bubble,
upon increasing the forcing
pressure, the bubble will eventually encounter the stable
equilibrium curve B;  tracking curve B on further increasing
the forcing pressure leads to an abrupt decrease in the bubble 
size. 
The experimentally observed abrupt decrease in the ambient radius
reflects the sharpness of the slope of curve B.
Eventually, at high enough forcing pressure, the bubble
will track curve C, the ambient radius now increasing
with forcing pressure.  

The most extensive experimental evidence in support of these
calculations
is
Holt and Gaitan's \cite{hol96} 
recent detailed measurement of phase diagrams in
the $R_0-P_a$ plane.
At low forcing pressure,
they observe bubble equilibria in agreement with classical calculations using
the pressure head ($p_\infty/P_0$=0.2) applied in the experiments. At very high
forcing pressure, the bubble equilibria agree with classical calculations,
but only when $p_\infty^{Ar}/P_0 = 0.002$ is used in the equations.
Between these two regimes, 
Holt and Gaitan find
the additional stable equilibria (curve B) of non-sonoluminescing bubbles
and an adjacent region 
(at $P_a \sim 1.2-1.3$atm) where bubbles dissolve, very similar to
that between curves B and C in Fig. \ref{r0vspa}.
In addition, Holt and Gaitan also find the the size of the dissolution
region decreases with increasing $p_\infty/P_0$, which also follows
directly from the theory.


To summarize,
the combination of the principles of sonochemistry and hydrodynamic
stability leads to a consistent picture of the stability of
sonoluminescing bubbles for gas mixtures and makes many predictions.
Our central statement is that it is only $p_\infty^{Ar}/P_0$ which
is relevant for SL stability in the high $P_a$ regime. We
included all available experimental data (with sufficient information
on $P_a$ and $p_\infty$) in our theoretical phase
diagram \cite{hil96} figure 1 to demonstrate the good agreement.
Finally, we suggest to measure the concentrations
of the reaction products as a function of time
for SBSL, as already done in
MBSL \cite{har87}. Nitrous acid production would
lead to a decrease in pH.
For an  estimate of an upper bound to the production rate we assume
that reactions at the collapse burn off all the nitrogen that diffuses
into the bubble during its expansion.
This amount of gas is estimated in ref.\ \cite{loe95} as $\Delta N_{N_2} =
2\pi D_{N_2} c_\infty^{N_2} R_m T /\mu_{N_2}$ per cycle. With typical
values of $R_m = 10 R_0$ for the maximal radius, $R_0 = 5\mu m$,
$D_{N_2}= 2\cdot 10^{-9} m^2/s$,
$c_\infty^{N_2} \approx 0.20 c_0^{N_2}$,
$c_0^{N_2}= 0.02 kg/m^3$, and $T=37\mu s$ one obtains 
$\Delta N_{N_2} \approx 3\cdot 10^{-18}$ mol per cycle or 
$\sim 3\cdot 10^{-10}$ mol of N$_2$ per hour converted to reaction
products. This results in a small but detectable pH decrease.


\noindent
{\bf Acknowledgments:}
We gratefully thank B.\ Barber, S.\ Grossmann,
L.\ Kadanoff, D.\ Oxtoby, and R. Holt for stimulating
discussions. 
This work has been supported by the DFG through its
SFB185,  by DOE, by NSF, and by the Chicago MRSEC program. 


\begin{thebibliography}{10}

\bibitem{cru94}
For recent reviews and further references, see
S.~J. Putterman, Scientific American {\bf 272},  32  (1995)
and L.~A. Crum, Physics Today {\bf 47},  22  (1994).


\bibitem{hil94}
R. Hiller, K. Weninger, S.~J. Putterman, and B.~P. Barber, Science {\bf 266},
  248  (1994).


\bibitem{sch36}
H. Schultes and H. Gohr, Angew. Chemie {\bf 49},  420  (1936).

\bibitem{ver88}
{\em Sonoluminescence in Ultrasound: Its
  chemical, physical and biological effects}, edited by K. Suslick (VCH,
  Weinheim, 1988).

\bibitem{hil96}
S. Hilgenfeldt, D. Lohse, and M.~P. Brenner, Phys. Fluids {\bf 8},  2808
  (1996);
M. Brenner, D. Lohse, and T. Dupont, Phys. Rev. Lett. {\bf 75},  954  (1995);
M. Brenner, D. Lohse, D. Oxtoby, and T. Dupont, Phys. Rev. Lett. {\bf 76},
  1158  (1996).

\bibitem{bar95}
B.~P. Barber, K. Weninger, R. L\"ofstedt, and S.~J. Putterman, Phys. Rev. Lett.
  {\bf 74},  5276  (1995).

\bibitem{loe95}
R. L\"ofstedt, K. Weninger, S.~J. Putterman, and B.~P. Barber, Phys. Rev. E
  {\bf 51},  4400  (1995).

\bibitem{gai90}
D.~F. Gaitan, Ph.D. thesis, The University of Mississippi, 1990.

\bibitem{mar85}
For reviews and further references, see
M.~A. Margulis, Ultrasonics  157  (1985);
K.~S. Suslick, Science {\bf 247},  1439  (1990).

\bibitem{ber95}
L. Bernstein and M. Zakin, J. Phys. Chem. {\bf 99},  14619  (1995).

\bibitem{hil92}
R. Hiller, S.~J. Putterman, and B.~P. Barber, Phys. Rev. Lett. {\bf 69},  1182
  (1992).

\bibitem{loe93}
R. L\"ofstedt, B.~P. Barber, and S.~J. Putterman, Phys. Fluids A {\bf 5},  2911
   (1993).

\bibitem{bar94}
B.~P. Barber {\it et~al.}, Phys. Rev. Lett. {\bf 72},  1380  (1994).

\bibitem{ber96}
L. Bernstein, M. Zakin, E. Flint, and K. Suslick, J. Phys. Chem. {\bf 100},
  6612  (1996).

\bibitem{fyr94}
M.~M. Fyrillas and A.~J. Szeri, J. Fluid Mech. {\bf 277},  381  (1994).

\bibitem{ple77}
A. Prosperetti,
J. Acoust. Soc. Am. {\bf 61},  17 (1977).

\bibitem{ell69}
A. Eller, J. Acoust. Soc. Am. {\bf 46},  1246  (1969);
A. Eller and H. G. Flynn, J. Acoust. Soc. Am. {\bf 37},  493  (1964).

\bibitem{hol96}
G. Holt and F. Gaitan, Phys. Rev. Lett. {\bf ??},  ???  (1996).

\bibitem{har87}
E. Hart, C.~H. Fischer, and A. Henglein, J. Phys. Chem. {\bf 91},  4166
  (1987).

\end{thebibliography}

\end{document}